\documentstyle[pre,aps,multicol,eqsecnum,epsf,rotate]{revtex}

\begin{document}
\draft
\tighten

\title{\LARGE Defect--induced condensation and central peak \\
              at elastic phase transitions}

\author{M.~Bulenda, F.~Schwabl, and U.C.~T\"auber $^*$}

\address{Institut f\"ur Theoretische Physik, \\
         Physik--Department der Technischen Universit\"at M\"unchen, \\
         James--Franck--Stra{\ss}e, D--85747 Garching, Germany}

\date{\today}
\maketitle

\begin{abstract}

Static and dynamical properties of elastic phase transitions under the
influence of short--range defects, which locally increase the transition
temperature, are investigated. Our approach is based on a Ginzburg--Landau
theory for three--dimensional crystals with one--, two-- or three--dimensional
soft sectors, respectively. Systems with a finite concentration $n_{\rm D}$ of
quenched, randomly placed defects display a phase transition at a temperature
$T_c(n_{\rm D})$, which can be considerably above the transition temperature
$T_c^0$ of the pure system. The phonon correlation function is calculated in
single--site approximation. For $T>T_c(n_{\rm D})$ a dynamical central peak
appears; upon approaching $T_c(n_{\rm D})$, its height diverges and its width
vanishes. Using an appropriate self--consistent method, we calculate the
spatially inhomogeneous order parameter, the free energy and the specific heat,
as well as the dynamical correlation function in the ordered phase. The
dynamical central peak disappears again as the temperatur is lowered below
$T_c(n_{\rm D})$. The inhomogeneous order parameter causes a static central
peak in the scattering cross section, with a finite $k$ width depending on the
orientation of the external wave vector ${\bf k}$ relative to the soft sector.
The jump in the specific heat at the transition temperatur of the pure system
is smeared out by the influence of the defects, leading to a distinct maximum
instead. In addition, there emerges a tiny discontinuity of the specific heat
at $T_c(n_{\rm D})$. We also discuss the range of validity of the mean--field
approach, and provide a more realistic estimate for the transition temperature.

\end{abstract}


\begin{multicols}{2}

\section{Introduction}

The influence of defects on the statics and dynamics of structural phase
transitions has been of considerable theoretical interest over the past two
decades \cite{Sch74,Hal76,Sch77,Tho77,Sch82,Sie84,Sch90,Sch91}. Especially the
appearence of a narrow central peak in the neutron scattering cross section,
well above the transition temperature, for both distortive \cite{Ris71} and
elastic structural transitions \cite{Shi71}, prompted various theoretical
studies dealing with local ordering phenomena around short--range static
defects (for a review of the experimental facts, see Ref.~\cite{Mul79} and
Ref.~\cite{Fle82}, and for a review of some theoretical results, see
Ref.~\cite{Cow80}).

E.g., in Ref.~\cite{Sch77} a one--dimensional model for continuous distortive
structural transitions was studied, with the order parameter coupling to a
single defect ($N_{\rm D}=1$). If the impurity locally increases the transition
temperature $T_c^0$ of the pure system, this leads to a local condensation of
the order parameter in the defect vicinity. In higher dimensions, for this
local order parameter condensation to occur, the defect potential strength must
exceed a certain minimal threshold. Such locally ordered regions in the
material emerging well above the pure transition temperature $T_c^0$ have
played a prominent role in some of the theories attempting to explain the
central peak phenomenon for distortive and elastic structural phase transitions
\cite{Sch74,Hal76,Sch77,Tho77,Sch82,Sie84}. In this paper, we extend previous
work on second--order ferroelastic phase transitions in $d=1$ \cite{Sch90} to
higher space dimensions $d$, taking into account the crystalline anisotropy. To
this end, we shall generalize the methods developed for the distortive case
\cite{Sch91} to (anisotropic) elastic systems, thus treating consistently a
random impurity system with finite defect concentration $n_{\rm D}$ (in the
thermodynamic limit, both the number of lattice sites $N \to \infty$ and the
number of defects $N_{\rm D} \to \infty$, but
$n_{\rm D} = N_{\rm D}/N = {\rm const.}$).

In the framework of our mean--field approach, we shall find that defects which
locally soften the crystal may induce a true phase transition at a temperature
$T_c(n_{\rm D}) > T_c^0$. Below this defect--induced phase transition
temperature a spatially inhomogeneous order parameter emerges, whose average
value remains very small in the vicinity of $T_c(n_{\rm D})$ and only becomes
noticeable near $T_c^0$. Similarly, thermodynamic quantities (static
susceptibility, specific heat, etc.) display prominent, broadened maxima near
$T_c^0$ suggesting a "rounded" phase transition; however, the "true"
singularities occur at $T_c(n_{\rm D})$, but may not be seen in experiment at
all, as their amplitude is only proportional to the defect concentration
$n_{\rm D}$. The Bragg peaks of the low--temperature phase already appear in
the scattering cross section for $T < T_c(n_{\rm D})$; as a consequence of the
spatial inhomogeneity of the order parameter, they are accompanied by elastic
Huang scattering peaks with finite $q$ width. Furthermore, very close to
$T_c(n_{\rm D})$ an additional dynamic central peak emerges, which may be
interpreted as a dynamical precursor to the defect--induced phase transition.

These mean--field results of course neglect order parameter fluctuations, and
exaggerate cooperative behavior. In reality, at $T \approx T_c(n_{\rm D})$
localized order parameter clusters appear, whose orientations however strongly
fluctuate in space. Only at a lower temperature $T_{\rm ord} < T_c(n_{\rm D})$
(if at all) will they form a collective state with uniform orientation, i.e.:
the spatially inhomogeneous configuration predicted by mean--field theory. In
order to provide a more realistic estimate of the proposed defect--induced
transition temperature, we consider the cluster orientations as effectively
Ising--like degrees of freedom, and then determine the cluster ordering
temperature $T_{\rm ord}$ by calculating the free--energy difference of states
with parallel and opposite orientation, respectively. Thus the onset of the
order parameter, the Bragg peaks, and the Huang scattering will be shifted to
somewhat lower temperatures, and the results of this work can essentially be
used if $T_c(n_{\rm D})$ is replaced by $T_{\rm ord}$. Provided that
$T_{\rm ord}$ is still considerably larger than $T_c^0$, we thus expect the
behavior of the thermodynamic quantities near $T_c^0$ to be very similar to the
results presented here.
 
This paper is organized as follows: In Sec.~II we introduce the
Ginzburg--Landau functional for a $d$--dimensional system, with one
$m$--dimensional soft sector, including randomly distributed point defects. The
corresponding Langevin--type equation of motion is formulated. Furthermore we
present an expression for the density--density correlation function, which
serves as a starting point for subsequent considerations. In Sec.~III the
phonon response function is evaluated in the high--temperature phase, and the
emergence of a dynamical central peak and a defect--induced phase transition
well above $T_c^0$ is demonstrated. In Sec.~IV we proceed to the ordered
low--temperature phase, by using a suitable self--consistent approach. We
determine the spatially inhomogeneous order parameter, the free energy and
specific heat, as well as the phonon correlation function, and discuss the
singularities in these quantities. In addition, the scattering cross section
$S({\bf k})$ (i.e., the density--density correlation function) is studied. In
Sec.~V we leave the realm of mean--field theory, and provide an estimate of the
``true'' defect--induced transition temperature (for the isotropic case), by
identifying it with that temperature where already existing, but still
fluctuating clusters condense to form a non--zero average order parameter. In
Sec.~VI, we briefly discuss the case of extended disorder (line or planar
defects), and in Sec.~VII we finally summarize and discuss our results.

\section{General Equations}

\subsection{Model}

In order to describe elastic phase transitions of second order in $d$
dimensions with an $m$--dimensional soft sector, we use an expansion of the
elastic free energy of the unperturbed crystal with respect to phonon
normal coordinates $Q_{{\bf k}}$, \cite{Fol76,Cow76}.
We disregard
non--critical polarizations; furthermore, aiming at the long--wavelength limit
we keep 
only the lowest--order terms in the wavevector expansion of the dispersion
relation of the acoustic phonons. The wavevector ${\bf k}$ is then
decomposed into its 
$m$--dimensional "soft" components ${\bf p}$, and its $(d-m)$--dimensional
"stiff" part ${\bf q}$, respectively: ${\bf k}=({\bf p},{\bf q})$. Folk, Iro,
and Schwabl have shown that terms of the form $q^4$ or $q^2 p^2$ are irrelevant
(in the renormalization group sense) and do not affect the critical behavior of
the system \cite{Fol76}.
In this spirit we use the following effective  free energy
\begin{eqnarray}
 F &=& \int d^dk \int d^dk' \frac{1}{2} \left[
      \left( a p^2 + b q^2 + c p^4 \right) \delta({\bf k}-{\bf k}')
                Q_{{\bf k}} Q_{-{\bf k}'} \right] \nonumber\\
&&+ {\cal O}(Q_{{\bf k}}^4) \ .
\label{Hupsi}
\end{eqnarray}
The coefficient $a$ is assumed to depend linearly on temperature, vanishing at
$T_c^0$: $a=a'(T-T_c^0)$; the very weak temperature dependence of the
Ginzburg--Landau coefficients $b$ and $c$ is neglected.

In order to describe the influence of short--range defects, which locally
increase the transition temperature, we assume that each defect creates a
short--range potential at its site, thus locally modifying the coefficients $a$
and $b$ of the Ginzburg--Landau functional (\ref{Hupsi}); being interested in
long--wavelength properties of the system, we can thus model the defect
potential in the continuum by a $\delta$ function. The coefficient $a$ will be
particularly sensitive to such a modification, as it becomes very small near
the transition. For the coefficient $c$ and the higher--order coefficients the
defect influence is less important and will be neglected. We thus arrive at the
following Ginzburg--Landau functional for the perturbed system,
\begin{eqnarray}
 F &=& \int d^dk \int d^dk' \frac{1}{2} \left[
     \left( a p^2 + b q^2 + c p^4 \right) \delta({\bf k}-{\bf k}')
                                                    Q_{{\bf k}}
                                                    Q_{-{\bf k}'}\right.\nonumber\\ 
     && \left.-
     \phi_{{\bf k},{\bf k}'} {\bf k} {\bf k}' Q_{{\bf k}} Q_{-{\bf k}'} \right]
     + {\cal O}(Q_{{\bf k}}^4) \ ,
     \label{Fritzi}
\end{eqnarray}
where $\phi_{{\bf k},{\bf k}'}$ denotes the Fourier transform of the impurity
potential (created by $N_{\rm D}$ defects)
\begin{equation}
\label{defpot}
 \phi({\bf r}) = U \sum_{i_{\rm D}=1}^{N_{\rm D}}
                   \delta({\bf r}-{\bf r}_{i_{\rm D}}) \ .
\end{equation}
The defect strength $U = a_0^d \lambda$ is taken to be positive, and therefore
the transition temperature is locally increased at the impurities (here, $V$ is
the volume of the system, and $a_0^d = V/N$ denotes the volume of the unit
cell).

The dynamics of the elastic crystal are governed by a Langevin--type equation 
of motion for the soft acoustic phonons \cite{Fol79},
\begin{equation}
\label{Babs}
 M \omega^2 Q_{{\bf k}} = -\frac{\delta F}{\delta Q_{-{\bf k}}}
                          - i M \omega (D p^2 + \tilde{D} q^2) Q_{{\bf k}}
                          + r_{{\bf k}} + h_{{\bf k}} \ .
\end{equation}
The term on the left--hand side of Eq.(\ref{Babs}) describes the acceleration,
while the first term on the right--hand side provides the restoring force
driving the system towards its equilibrium configuration. Note that we have
introduced two different diffusive damping constants $D$ and $\tilde{D}$ for
the soft and stiff sectors, respectively. $r_{{\bf  k}}$ denotes a stochastic
force with vanishing average, $\langle r_{{\bf  k}} \rangle = 0$; its second
moment satisfies an Einstein relation, guaranteeing that $\exp(-F/k_{\rm B}T)$
is the equilibrium probability distribution. Finally, $h$ is an external field
which couples linearly to the order parameter. Eq.(\ref{Babs}) will be the
basis for our discussion of the dynamical properties in the subsequent
chapters.

\subsection{Density--density correlation function}

In the following we shall primarily use a discrete lattice representation of
the elastic system under consideration. The dynamic structure factor observed
in scattering experiments is related to the Fourier--transformed
density--density correlation function. Denoting the thermodynamical average by 
$\langle \ldots \rangle$, its definition is
\begin{equation}
\label{Scott}
 S({\bf k},\omega) = \int dt e^{i \omega t} \bigg\langle \frac{1}{N}
  \sum_{1 \le i,j \le N} e^{-i {\bf k} \left[ {\bf a}_i + {\bf u}_i(t)
  \right]} 
         e^{i {\bf k} \left[ {\bf a}_j + {\bf {u}}_j(0) \right]} \bigg\rangle
         \ , 
\end{equation}
where ${\bf a}_i$ denote the Bravais lattice sites (of the high--temperature
phase), and ${\bf u}_i$ the displacements from these equilibrium positions.

In the discrete representation, with $N$ lattice sites, we can write the
Fourier--transformed defect potential as
\begin{equation}
\label{Stupsi}
 \phi_{{\bf k}{\bf k}'} = \frac{1}{N} \sum_{i,j=1}^N \left(
        \sum_{i_{\rm D}=1}^{N_{\rm D}} \lambda \delta_{i,i_{\rm D}} \delta_{ij}
                       \right) e^{-i({\bf k} {\bf x}_i-{\bf k}' {\bf x}_j)}
                       \ . 
\end{equation}
In a system with quenched, randomly distributed defects, all physical
quantities have to be averaged over all possible defect configurations
\cite{Lan61}. We denote this configurational average by
$\langle \langle \ldots \rangle \rangle$; its formal definition reads
\begin{equation}
\label{Grump}
 \langle \langle \ldots \rangle \rangle = \prod_{j=1}^{N_{\rm D}}
                   \left[ \frac{1}{N} \sum_{i_{{\rm D}_j}}^N \right] \ldots \ .
\end{equation}

In order to evaluate $\langle\langle S({\bf k},\omega)\rangle\rangle$, we
introduce a cumulant expansion for the combined thermal and configurational
averages of $e^{i{\bf k}[{\bf u}_j(0)-{\bf u}_i(t)]}$ and keep the terms up to
second order. Next we decompose the deviations ${\bf u}_i(t)$ into a static
contribution $\boldmath{\psi}_i$ and a fluctuating part ${\bf v}_i(t)$, and
expand 
the exponential. Eventually one arrives at the following formula for the
dynamical structure factor (for more details on the derivation, see
Ref.~\cite{Sch91})
\begin{eqnarray}
 \langle \langle S({\bf k},\omega) \rangle \rangle &=&
  \left[ N
 \sum_{{\bf g}} \delta_{{\bf k},{\bf g}} + \sum_{\alpha \beta} k^\alpha k^\beta
    \langle \langle S_c^{\alpha \beta}({\bf k}) \rangle \rangle \right] 
  \times \nonumber\\
  && e^{-2W} 2 \pi \delta(\omega) + \left[ \sum_{\alpha \beta} k^\alpha
  k^\beta 
             D^{\alpha \beta}({\bf k},\omega) \right] e^{-2W} \ . \nonumber\\
 \label{Struc}
\end{eqnarray}
The three different contributions to the dynamical structure factor in
Eq.(\ref{Struc}) are (i) the elastic Bragg peaks appearing at the reciprocal
lattice vectors ${\bf g}$ of the actual crystal structure, given by the
condition
\begin{equation}
\label{recip}
 e^{i {\bf g} \left( {\bf a}_i + \langle \langle \mbox{$\boldmath{\psi}_i
 $}\rangle \rangle \right)} = 1 \ ;
\end{equation}
(ii) an additional static contribution to the structure factor arising from
elastic scattering from the random variations of the local order parameter
(Huang scattering)
\begin{equation}
\label{Huang}
 S_c^{\alpha \beta}({\bf k}) = \frac{1}{N} \sum_{i,j}
                              e^{-i {\bf k} \left( {\bf a}_i-{\bf a}_j \right)}
       \left( \psi_i^\alpha \psi_j^\beta - \langle \langle \psi_i^\alpha
       \rangle \rangle \langle \langle \psi_j^\beta \rangle \rangle \right) \ ;
\end{equation}
and (iii) the dynamical phonon--phonon correlation function
\begin{equation}
\label{corfun}
 D^{\alpha \beta}({\bf k},\omega) = \int dt e^{i\omega t}
   \bigg\langle \bigg\langle \frac{1}{N} \sum_{i,j}
                              e^{-i {\bf k} \left( {\bf a}_i-{\bf a}_j
                              \right)} 
       \langle v_i^\alpha(t) v_j^\beta(0) \rangle \bigg\rangle \bigg\rangle \ ,
\end{equation}
which is connected with the dynamic phonon response function via the
(classical) fluctuation--dissipation theorem
\begin{equation}
\label{FDT}
 D^{\alpha \beta}({\bf k},\omega) = \frac{2k_{\rm B}T}{\omega}
                                \mbox{Im}\ G^{\alpha \beta}({\bf k},\omega) \ .
\end{equation}
Finally,
\begin{eqnarray}
 W &=& \frac{1}{2} \sum_{\alpha \beta} k^\alpha k^\beta \left[
     \bigg\langle \bigg\langle \left( \psi_i^\alpha -
                          \langle \langle \psi_i^\alpha \rangle \rangle \right)
     \left( \psi_i^\beta - \langle \langle \psi_i^\beta \rangle \rangle \right)
                          \bigg\rangle \bigg\rangle \right.
                          \nonumber\\
&& \left. + \bigg\langle \bigg\langle
         \langle v^\alpha_i v^\beta_i \rangle \bigg\rangle \bigg\rangle \right]
\end{eqnarray}
is the Debye--Waller factor. Eq.(\ref{Struc}) may be used for elastic as well
as for distortive phase transitions. For antiferrodistortive transitions one
has to sum over the distinct sublattices in addition (see Ref.~\cite{Sch91}).
The dynamic phonon--phonon correlation function (\ref{corfun}) and the static
Huang scattering contribution (\ref{Huang}) will be discussed in more detail
below.

\section{High--temperature phase}

In order to calculate the phonon correlation function in the high temperature
phase, Eq.(\ref{Fritzi}) is inserted in the equation of motion (\ref{Babs}). 
Note that nonlinearities in the phonon normal coordinates are neglected, and
thus fluctuations are only being accounted for in the Gaussian approximation.
Upon differentiating the resulting expression with respect to $h_{{\bf k}'}$,
and transcribing it to the corresponding discrete version, and finally using 
the fact that the average of the stochastic force $r_{{\bf k}'}$ vanishes, one 
arrives at the following mean--field recursion relation for the phonon response
function
\begin{equation}
\label{Recur}
 G_{{\bf k}{\bf k}'} = G_{0{\bf k}} \delta_{{\bf k}{\bf k}'} +
        G_{0{\bf k}} \sum_{{\bf k}''} \phi_{{\bf k}{\bf k}''} {\bf k} {\bf k}''
                                                      G_{{\bf k}''{\bf k}'} \ ,
\end{equation}
with the free phonon propagator
\begin{equation}
 G_{0{\bf k}}^{-1} = - M \omega^2 + a p^2 + b q^2 + c p^4
                     - i M \omega (D p^2 + \tilde{D} q^2) \ .
\end{equation}
Eq.(\ref{Recur}) may then be systematically iterated, and the configurational
average (\ref{Grump}) performed; e.g., applying a standard diagrammatic
technique helps in collecting all the contributions to a certain order in the
defect potential strength $\lambda$ \cite{Lan61,Ell74}. The configurational
average yields a translationally invariant response function,
\begin{equation}
 \langle \langle G_{{\bf k}{\bf k}'} \rangle \rangle =
                                       G_{{\bf k}} \delta_{{\bf k}{\bf k}'} \ .
\end{equation}
Upon collecting all contributions which are proportional to the impurity
concentration $n_{\rm D} = N_{\rm D}/N$ (single--site approximation; for more
details, see Ref.~\cite{Sch91}), the result for the phonon susceptibility is
\begin{eqnarray}
 G_{{\bf k}}^{-1} &=& - M \omega^2 + a p^2 + b q^2 + c p^4
                    - i M \omega (D p^2 +\tilde{D} q^2)\nonumber\\
                    &&
           - \frac{\lambda k^2 n_{\rm D}}{1-\lambda (a_0 / 2 \pi)^d I_d(m)} \ .
\label{Petra}
\end{eqnarray}
Here we have introduced the abbreviation
\end{multicols}
\begin{eqnarray}
 I_d(m) &=& \int_{0}^{\Lambda} \frac{(p^2 + q^2)}
    {- M \omega^2 + a p^2 + b q^2 + c p^4 - i M \omega (D p^2 + \tilde{D} q^2)}
                                               d^mp \ d^{d-m}q \label{Karin} \\
 &=& m (d-m) \tau_m \tau_{d-m} \int_{0}^{\Lambda}
    \frac{(p^2 +q^2) p^{m-1} q^{d-m-1}}
      {-M \omega^2 + a p^2 + b q^2 + c p^4 - i M \omega (D p^2 +\tilde{D} q^2)}
dp \ dq \ , \nonumber
\end{eqnarray}
\begin{multicols}{2}
and $\tau_n$ is the volume of the $n$--dimensional unit sphere
\begin{equation}
\label{Tau}
 \tau_n = \frac{\pi^{n/2}}{\Gamma \left(\frac{n}{2}+1\right)} \ .
\end{equation}
$\Lambda$ denotes a natural short--wavelength cutoff (e.g., corresponding to 
the Brillouin zone boundary), which also helps to ensure the convergence of the
integral $I_d(m)$. Note that the dimension $m$ of the soft sector in $k$ space
explicitly enters in Eq.(\ref{Karin}), and thus determines the importance of
the fluctuation contributions.

Eq.(\ref{Petra}) implies the very remarkable result that due to the coupling to
the softening defects, the entire system may become unstable towards a new 
ground state with finite average order parameter at a certain temperature 
$T_c(n_{\rm D})$, depending on the defect concentration $n_{\rm D}$; the
criterion for this instability is
\begin{equation}
\label{instab}
 \lim_{k \to 0} \left[ G_{{\bf k}}^{-1}(\omega=0) / k^2 \right] = 0 \ .
\end{equation}
As in the distortive case \cite{Sch91}, we find that in general a certain
minimal defect strength is required for this instability to occur; yet, once
this defect--induced phase transition does exist, the associated transition
temperature $T_c(n_{\rm D})$ can be considerably higher than that of the pure 
system, $T_c^0$. (In Sec.~V, we shall comment on the validity of the 
mean--field approach, and estimate the transition temperature on a more
realistic basis.)

We have investigated three--dimensional systems with one one--, two-- or
three--dimensional soft sector, respectively. The qualitative features were 
found to be very similar in all these cases. The following figures refer to a 
three--dimensional system with a single one--dimensional soft sector. We have
tried to use model parameters appropriate for ${\rm Nb_3Sn}$, which displays a
second--order elastic phase transition near $T = 45 {\rm K}$ \cite{Reh68}; 
accordingly, we have used numerical values calculated from
Refs.~\cite{Axe73,Reh68} (Table I). Thus, we have taken $T_c^0 = 45 {\rm K}$,
$n_{\rm D} = 10^{-5}$, and adjusted the defect strength in order that
$T_c(n_{\rm D}) = 65 K$. However, a few remarks are in place here to explain
some sources of inaccuracies. The assumption that the Ginzburg--Landau
parameter $a$ is merely linearly temperature--dependent is valid only near the
phase transition temperature of the pure system $T_c^0$. Furthermore we
approximated $c$ and $b$ as independent of temperature, and in addition assumed
$b$ to be independent of the direction of the ${\bf k}$ vector in the stiff
plane. This is not generally the case for ${\rm Nb_3Sn}$, but appears to
describe the critical region well. The numerical values of the diffusion
constant $D = \tilde{D}$ and the nonlinearity $d$ (see below) had to be
estimated without reference to any experiment.

Fig.1 depicts the phonon correlation function $D({\bf k},\omega)$
[Eq.(\ref{FDT})] for different temperatures $T > T_c(n_{\rm D})$, evaluated for
several angles $\theta$ between the external wave vector and the soft sector.
As becomes apparent in Fig.1, a dynamical central peak in the phonon
correlation function emerges in addition to the soft phonon peak (compare
Ref.\cite{Sch90} for the one--dimensional case). The height of the central peak
grows, and its width decreases as $T_c(n_{\rm D})$ is approached. The intensity
of the central peak decreases upon increasing the angle $\theta$ between the
wave vector ${\bf k}$ and the soft sector. This reflects the fact that
wavevectors in the stiff sector do not probe the critical properties of the
material. The dynamical central peak may thus be understood as a dynamic
precursor to the defect--induced second--order phase transition at
$T_c(n_{\rm D})$.

\section{Low--temperature phase}

In this section, we use a self--consistent approach designed for the
calculation of the order parameter $\nu$, the specific heat, the phonon
correlation function and finally the dynamical structure factor in the ordered
phase, i.e., for $T < T_c(n_{\rm D})$.

\subsection{Order parameter}

The starting point for the calculation of the order parameter is the full 
nonlinear Ginzburg--Landau functional, which in the discrete lattice
representation reads
\begin{eqnarray}
 F &=& \frac{1}{2} \sum_{i,j=1}^N \nu_i G_{0ij}^{-1} \nu_j -
     \frac{\lambda}{2} \sum_{i=1}^N \sum_{i_{\rm D}=1}^{N_{\rm D}} \nu_i^2
                                                          \delta_{ii_{\rm D}} +
      \frac{d}{4} \sum_{i=1}^N \nu_i^4 \nonumber\\
&&- \sum_{i=1}^N h_i \nu_i \ .
\label{Ginzi}
\end{eqnarray}
Here $\nu_i$ denotes the value at lattice site $i$ of that combination of 
strain tensor components serving as the order parameter for the transition,
$h_i$ is the corresponding external stress acting on site $i$, and the static
propagator $G_{0ij}$ is defined by its Fourier transform
\begin{eqnarray}
 G_{0{\bf k}}^{-1} &=& a+ck^2 \quad \mbox{for $d=m$,} \nonumber \\
 G_{0{\bf k}}^{-1} &=& \frac{a p^2 + b q^2 + cp^4}{k^2} \quad
                                               \mbox{for $d>m$.} \label{static}
\end{eqnarray}
In the framework of the Ginzburg--Landau approximation, i.e., neglecting order
parameter fluctuations, the following stationarity condition can be derived
(with $h_i = h = {\rm const.}$)
\begin{equation}
\label{Stats}
 \frac{\delta F}{\delta \nu_i} = 0 \quad \Leftrightarrow \quad
                         \sum_{j} G_{0ij}^{-1} \nu_j - \lambda \sum_{i_{\rm D}}
                                 \nu_i \delta_{i,i_{\rm D}} + d \nu_i^3 = h \ .
\end{equation}

A general solution of Eq.(\ref{Stats}), with its combined nonlinearity and
randomness, poses a difficult problem. We thus use an additional approximation,
namely the following ansatz \cite{Sch91} for the thermodynamical average of the
order parameter (denoted by $\bar{\nu}$),
\begin{equation}
\label{Deppi}
 \bar{\nu}_i = A + B \sum_{i_{\rm D}} \delta_{i,i_{\rm D}} \ , 
\end{equation}
i.e.: we assume that the order parameter at each lattice point $i$ may be
written as the sum of a homogeneous background $A$ and an additional 
contribution $B$, if there is a defect at site $i$, thus enhancing the total
value of the order parameter to $A+B$ at the defect sites. Thus we explicitly
assume that at all defect sites the order parameter points in the same
direction, and in addition neglect the spatial variation of the order parameter
near the defects. However, as we shall see shortly, the second, seemingly very
crude approximation already contains the possible relevant modifications caused
by the impurities, namely (i) an enhancement of the spatially averaged order
parameter (corresponding to the parameter $A$), and (ii) the ensuing
``screening'' of the defect potential (described by the coefficient $B$). The
more stringent approximation is the uniform orientation of the defect clusters,
as implied by the mean--field approach (see Sec.~V).

Inserting Eq.~(\ref{Deppi}) into the stationarity equation (\ref{Stats}) yields
the recursion relation
\begin{equation}
\label{Sabine}
 \bar{\nu}_{{\bf k}} = h \delta_{{\bf k}0} \tilde{G}_0({\bf k}) +
       \tilde{G}_0({\bf k}) \sum_{{\bf k}'} \tilde{\phi}_{{\bf k}{\bf k}'}
                                                       \bar{\nu}_{{\bf k}'} \ ,
\end{equation}
where we have introduced a renormalized propagator
\begin{equation}
\label{renpro}
 \tilde{G}_0({\bf k})^{-1} = G_0({\bf k})^{-1} + d A^2 \ ,
\end{equation}
and a screened defect potential $\tilde{\phi}_{{\bf k}{\bf k}'}$ with
weakened strength [see Eq.(\ref{Stupsi})]
\begin{equation}
 \tilde{\lambda} = \lambda - d [(A+B)^2 - A^2] \ .
\end{equation}
From Eq.(\ref{Sabine}) and the averaged stationarity equation we may derive two
coupled nonlinear equations that uniquely determine the mean order parameter:
(i) Iterating the recursion relation (\ref{Sabine}) in a similar way as for the
dynamics in the previous paragraph, performing the configurational average, and
summing the single--site contributions, one arrives at
\begin{equation}
\label{Rita}
 \frac{\langle \langle \bar{\nu} \rangle \rangle}{h} = \left[ a + d A^2 -
 \frac{\tilde{\lambda} n_{\rm D}}{1-\tilde{\lambda} (a_0 / 2 \pi)^d J_d}
                                                               \right]^{-1} \ ,
\end{equation}
with the abbreviation
\begin{equation}
\label{Integral}
 J_d = \int d^dk \frac{k^2}{(a + d A^2) p^2 + (b + d A^2) q^2 + c p^4} \ .
\end{equation}
(ii) On the other hand, immediate averaging of Eq.(\ref{Sabine}) yields
\begin{equation}
\label{Pia}
 (a + d A^2) (A + n_{\rm D} B) - \tilde{\lambda} n_{\rm D} (A+B) - h = 0 \ .
\end{equation}
Very assuringly, Eqs.(\ref{Rita}) and (\ref{Pia}) yield non--zero solutions for
$\langle \langle {\bar \nu} \rangle \rangle$ precisely below $T_c(n_{\rm D})$ 
as determined from the high--temperature phase. Fig.2 shows that the order 
parameter of the perturbed system as function of $T$ looks similar to the 
corresponding curve for the pure system, with the singularity at $T_c^0$ being
smeared out by the disorder. The order parameter sets in continuously at 
$T_c(n_{\rm D})$, with the usual mean--field exponent $\beta = 1/2$, assumes 
small but finite values in the range $T_c(n_{\rm D}) > T > T_c^0$, and starts 
to grow to larger values only in the vicinity of $T_c^0$. Thus the transition 
temperature of the pure system remains an important parameter even in the 
perturbed system, while the true phase transition at $T_c(n_{\rm D})$ may in
fact be hardly noticeable in experiments. As before, the results for a 
three--dimensional system with one one--dimensional soft sector are depicted, 
but the qualitative features remain essentially the same in the cases of a
two-- or three--dimensional soft sector.

\subsection{Specific heat}

From the knowledge of the mean order parameter, we can readily calculate the 
(averaged) free energy from Eq.(\ref{Ginzi}) in Landau approximation, and via
\begin{equation}
\label{Spezi}
 C_v = - T \left( \frac{\partial ^2 F}{\partial T^2} \right)_V
\end{equation}
derive the specific heat $C_v$, see Fig.3. Obviously, the discontinuity at
$T_c^0$ has been smeared out, in place of which a tiny jump emerges at 
$T_c(n_{\rm D})$. Although the phase transition clearly occurs at 
$T_c(n_{\rm D})$, the transition temperature $T_c^0$ of the pure system remains
of considerable importance; e.g., there is a distinct maximum of the specific 
heat near $T_c^0$, while the extremely minute jump at $T_c(n_{\rm D})$ might 
not be experimentally detectable at all.

\subsection{Phonon correlation function in the ordered phase}

In order to find the phonon correlation function in the temperature region with
a finite order parameter, one again has to start from the full Ginzburg--Landau
functional and use the ansatz for the order parameter (\ref{Deppi}). The 
crucial point is that one may then absorb the nonlinear term of the equation of
motion in modified coefficients of the linear terms as follows
\begin{equation}
\label{Renorm}
 a \rightarrow a + 3 d A^2 \, , \quad
 b \rightarrow b + 3 d A^2 \, , \quad
 \lambda \rightarrow \lambda - 3 d B (2 A + B) \\ .
\end{equation}
With these modifications one can use the same equations as in the
high--temperature phase.

The result is depicted in Fig.4. The central peak in the correlation function
disappears again when the temperature is lowered below $T_c(n_{\rm D})$. This
dynamical central peak is thus confined to the region around $T_c(n_{\rm D})$.
As in the high--temperature phase, the intensity of the central peak decreases
upon increasing the angle between the external ${\bf k}$ vector and the soft
sector (with fixed temperature).

In Fig.5 the static phonon susceptibility $G({\bf k}) = G_{{\bf k}}(\omega=0)$
(i.e., the inverse elastic constant, as modified by the defects) is shown.
The small but sharp peak at $T_c(n_{\rm D})$ reflects the preordering of the
defect regions, while the broad and much more prominent peak near $T_c^0$
corresponds to the ordering of the pure bulk crystal, though under the
influence of the randomly spaced fields originating from the defect clusters;
compare Figs.2 and 3.

\subsection{Dynamical structure factor}

In order to describe scattering experiments, we have to calculate the 
density--density correlation function of Sec.2. The first term in 
Eq.(\ref{Struc}) yields the Bragg scattering, and does not require any further
comment; the third term is connected with the phonon correlation function, and
has been discussed in the previous subsection. We therefore turn our attention 
to the second term. Taking into account the soft acoustic phonon mode only, as
above, we have to calculate the configurational average of
$k^2 \psi_{{\bf k}}\psi_{-{\bf k}}$ [Eq.(\ref{Huang})]; using the same
approximations as in the beginning of Sec.3, we may use equation (\ref{Sabine})
in the form
\begin{equation}
\label{Jippi}
 k \psi_{{\bf k}} = h \tilde{G}_0(0) \delta_{{\bf k}0} +
   \tilde{G}_0({\bf k}) \sum_{{\bf k}'} \tilde{\phi}_{{\bf k}{\bf k}'}
                                                      \bar{\nu}_{{\bf k}'} \ , 
\end{equation}
which yields
\begin{eqnarray}
 k^2 \psi_{{\bf k}} \psi_{-{\bf k}} &=& h^2 \tilde{G}_0(0)^2 \delta_{{\bf k}0} +
 h \tilde{G}_0(0)^2 \sum_{{\bf k}'} \tilde{\phi}_{0{\bf k}'}
  \bar{\nu}_{{\bf k}'} \delta_{{\bf k}0} + \nonumber\\
&&\tilde{G}_0({\bf k})\sum_{{\bf k}'}
   \tilde{\phi}_{{\bf k}{\bf k}'} \bar{\nu}_{{\bf k}'} \bar{\nu}_{-{\bf k}} \ .
\label{Juppi}
\end{eqnarray}
For this equation again a diagrammatic representation can be derived
\cite{Sch91}, and in single--site approximation (i.e.: to order $n_{\rm D}$) we
find the following result (${\bf k}_L$ denotes the components of the wave
vector which are parallel to the polarization of the soft mode)
\begin{eqnarray}
\label{Heidei}
 k^2 S_c^{LL}({\bf k}) &=& n_{\rm D} \tilde{\lambda}^2
          \langle \langle \bar{\nu} \rangle \rangle^2 (k^L)^2
          \tilde{G}_0({\bf k}) \tilde{G}_0(-{\bf k}) \times \nonumber\\ 
      &&  \left[ 1 - \frac{\tilde{\lambda}}{N} \sum_{k'} \tilde{G}_0({\bf k'})
                                                               \right]^{-2} \ .
\end{eqnarray}
This expression can be further reduced using Eq.~(\ref{Rita}). Finally, we 
arrive at
\begin{equation}
\label{Didudeldei}
 k^2 S_c^{LL}({\bf k}) = (a + d A^2)^2
       \frac{\langle \langle \nu \rangle \rangle^2}{n_{\rm D}} (k^L)^2
                                                     \tilde{G}_0({\bf k})^2 \ .
\end{equation}

Collecting all results, the final expression for the dynamical structure factor
reads
\end{multicols}
\begin{equation}
\label{Martina}
 \langle \langle S({\bf k},\omega) \rangle \rangle =
 \left[ N \sum_{{\bf g}} \delta_{{\bf g},{\bf k}} + \frac{(k^L)^2}{k^2}
 \frac{\langle \langle \bar{\nu} \rangle \rangle^2}{n_{\rm D}}
 \tilde{G}_0({\bf k})^2 (a + d A^2)^2 \right]  
e^{-2W} 2 \pi \delta(\omega)+ (k^L)^2 D({\bf k},\omega) e^{-2W} \ .
\end{equation}
\begin{multicols}{2}
Thus we have found three distinct effects, namely (i) new positions of the
Bragg peaks as a result of the finite order parameter [shifted and possibly new
reciprocal lattice vectors, see Eq.(\ref{recip})] ; (ii) Huang scattering as a
result of the spatially inhomogeneous order parameter configuration, leading to
a static central peak with finite width $\gamma = \sqrt{(a + d A^2)/c}$ (in the
soft sector) in Fourier space; and (iii) inelastic scattering, described by the
phonon correlation function. Fig.6 shows how the intensity of the Huang
scattering varies with temperature for different wave vectors ${\bf k}={\bf p}$
in the soft sector. This additional elastic contribution sets in at
$T_c(n_{\rm D})$, and then grows to considerable values near $T_c^0$.

\section{Estimate of the cluster ordering temperature}

All our previous results for the statics were based entirely on the 
Ginzburg--Landau approximation, and dynamic quantities were calculated in the 
Gaussian ensemble. This mean--field treatment of course neglects fluctuations,
and apart from the fact that the critical exponents will be changed near the
transition, we have to consider the possibility that the above described
defect--induced phase transition at $T_c(n_{\rm D})$ will disappear when 
fluctuations are properly taken into account. Namely, our mean--field approach
basically implies that as soon as local condensates form near the defects, they
immediately lock into some cooperative state and form a non--vanishing average 
order parameter. In reality, probably first these clusters may emerge at the 
defect positions, however still quite independently fluctuating between their 
different possible orientations. Only as the temperature is lowered even 
further, they will form a collective vibrational mode which finally condenses 
to a static order parameter at the ``true'' transition temperature 
$T_{\rm ord}$, with $T_c^0 \leq T_{\rm ord} \leq T_c(n_{\rm D})$, the 
mean--field transition temperature. One would expect that such collective 
behavior of the distinct localized order parameter clusters arises when the 
correlation length of the pure system $\xi$, which determines the size of the 
defect--induced condensates, becomes of the order of the average defect 
separation $r_{\rm D}$. A somewhat more favorable estimate results from the
argument that it should actually suffice when $\xi$ becomes large enough such
that the distinct condensates form a percolating cluster throughout the sample;
the condition for cooperative behavior then becomes
$\xi \approx (n_c)^{1/d} r_{\rm D}$, where $n_c$ denotes the percolation
threshold.

In the following we give a more precise estimate $T_{\rm ord}$, in order to see
if it may still be considerably above the transition temperature of the pure
system $T_c^0$. Our strategy is to calculate the free--energy difference
$\Delta F$ between the following two configurations in a two--defect system
below the temperature where localized clusters may form in two different
orientations: (i) both order parameter condensates oriented in the same
direction, and (ii) opposite condensate orientations. The ensemble of localized
clusters can then be effectively mapped onto an Ising system, with $\Delta F$
assuming the role of the exchange coupling. The critical temperature is now
readily estimated as $k_{\rm B} T_{\rm ord} \approx \Delta F / a_0^3$ ($a_0^3$
is the volume of the elementary cell). We emphasize that we shall restrict
ourselves to an isotropic system here, and consider the general case of an
order parameter $\varphi$ described by the usual Ginzburg--Landau expansion of
the free energy, which rather corresponds to the case of distortive structural
transitions, as studied in Ref.\cite{Sch91}. However, the qualitative behavior
is expected to be very similar for the anisotropic elastic phase transitions.

Using the continuum representation, the free energy for a system in three
dimensions with a single defect in the origin reads \cite{Sch77}
\begin{equation}
\label{freeen}
 F = \int d^3r \left( [a - \phi({\bf r})] \varphi({\bf r})^2 
   + c [\nabla \varphi({\bf r})]^2 + {d \over 2} \varphi({\bf r})^4 \right) \ ,
\end{equation}
where $\phi({\bf r})$ is the positive $\delta$ function defect potential with
strength $U = a_0^3 \lambda$. The stationarity equation then becomes
\begin{equation}
\label{statio}
 c \nabla^2 \varphi({\bf r}) = [a - \phi({\bf r})] \varphi({\bf r}) 
                                                     + d \varphi({\bf r})^3 \ ,
\end{equation}
which for $a > 0$ may be approximately solved by
\begin{eqnarray}
 &r > R: \qquad 
 &\varphi(r) \approx {\varphi_0 e^{r / \xi} \over 1 + r e^{2 r / \xi}/\xi} 
             \approx \varphi_0 {e^{-r / \xi} \over r / \xi} \\
 &r < R: \qquad
 &\varphi(r) \approx \varphi_0 \left( 1 - {3 r^2 \over 2 \xi^2} \right) \ ,
\end{eqnarray}
where $\xi = \sqrt{c/a}$ is the correlation length of the pure system for
$T > T_c^0$, $\varphi_0^2 \approx (4 \pi R^3 \lambda /3 - 10 a)/d$, and
$R^3 \lambda = 120 \pi c^3 (U_m^{-1} - U^{-1})^2$, with $U_m = 2 \pi c a_0$
denoting the minimum defect strength required for the local order parameter
condensation to occur.

Using these results, we can proceed towards the two--defect system with
$\phi({\bf r}) = U \delta(x) \delta(y) [\delta(z-r_{\rm D}/2) +
\delta(z+r_{\rm D}/2)]$ by a simple linear superposition ansatz; i.e., we shall
evaluate the free energy difference between the states
\begin{equation}
 \varphi_\pm = \varphi(x,y,z-r_{\rm D}/2) \pm \varphi(x,y,z+r_{\rm D}/2) \ .
\end{equation}
By inserting into Eq.~(\ref{freeen}) one readily finds the defect contribution
\begin{equation}
 \Delta F_{\rm D} \approx - 8 U \varphi_0^2 (\xi / r_{\rm D})
                                                         e^{-r_{\rm D}/\xi} \ ,
\end{equation}
as well as the linear overlap integral (conveniently evaluated using elliptical
coordinates)
\begin{equation}
 \Delta F_{\rm lin} \approx 16 \pi c \varphi_0^2 \xi e^{-r_{\rm D} / \xi} \ ;
\end{equation}
the nonlinear overlap integral turns out to be of order $e^{-2 r_{\rm D}/\xi}$
and can thus be neglected for $\xi \leq r_{\rm D}$, when compared to the
previous terms. Hence we find for the required free--energy difference
\begin{equation}
 \Delta F \approx 16 \pi c \varphi_0^2 \xi e^{-r_{\rm D}/\xi} 
                             \left( 1 - {U \over 2 \pi c r_{\rm D}} \right) \ ;
\end{equation}
using the above numerical values, we see that the defect contribution can in
fact be neglected here.

Hence we arrive at our final estimate for the cluster ordering temperature, 
which we identify with the ``true'' defect--induced transition temperature
\begin{equation}
 k_{\rm B} T_{\rm ord} \approx 16 \pi a \varphi_0^2 (\xi / a_0)^3 
                                                         e^{-r_{\rm D}/\xi} \ .
\end{equation}
This expression may be cast into a somewhat more explicit form by observing
that the average defect separation can be written as
$r_{\rm D} = a_0 n_{\rm D}^{-1/3}$; thus the required defect concentration for
the transition to occur at a certain value $T = T_{\rm ord}$ becomes
\begin{equation}
 n_{\rm D} = \left( {\xi_c \over a_0} 
                 \ln \left[ {16 \pi a_c \phi_0^2 \over k_{\rm B} T_{\rm ord}} 
                    \left( {\xi_c \over a_0} \right)^3 \right] \right)^{-3} \ ,
\end{equation}
from which $T_{\rm ord}$ as function of $n_{\rm D}$ may be inferred by
inversion [$\xi_c = \sqrt{c/a_c}$, $a_c = a' (T_{\rm ord} - T_c^0)$]. The
result is depicted in Fig.7. It can be seen that the calculated cluster
ordering temperature may indeed be considerably above the phase transition
temperature of the pure system $T_c^0=45 \mbox{K}$, however, much larger
impurity concentrations $n_{\rm D}$ are required than in the previous
mean--field analysis.

For the anisotropic elastic systems discussed in the bulk of this paper,
fluctuations will be even less important. We conclude this section with the
remark that the upper critical dimension as function of the dimension $m$ of
the soft sector was found to be \cite{Fol76}
\begin{equation}
 d_c(m) = 2 + {m \over 2} \ ;
\end{equation}
thus in three dimensions mean--field theory yields exact results for a system
with a one--dimensional soft sector, while for the case of a two--dimensional
soft sector merely logarithmic corrections are to be expected.

\section{Extended disorder: line and planar defects}

We now return to the case of elastic phase transitions, and address the
question of the influence of extended defects in contrast to the previously
treated point disorder. Our system now contains randomly placed, but parallel
linear or planar defects; the accordingly modified correlated defect potential
(compare Eq.~\ref{defpot}), reads in the case of line disorder
\begin{equation}
   \phi({\bf r}) = U \sum_{i_{\rm D}=1}^{N_{\rm D}}
                   \delta(x-x_{i_{\rm D}}) \delta(y-y_{i_{\rm D}}) \ ,
\end{equation}
where $x,y,z$ are the components of ${\bf r}$, and $z$ denotes the direction 
parallel to the lines; $i_{\rm D}$ labels the $N_{\rm D}$ line defects. The
defect potential for planar defects is defined analogously, namely for planes
normal to the $x$ direction
\begin{equation}
   \phi({\bf r}) = U \sum_{i_{\rm D}=1}^{N_{\rm D}} \delta(x-x_{i_{\rm D}}) \ ,
\end{equation}
With these definitions the same calculations as before may be performed, and it
becomes obvious that the former integrals reduce to integrals over the
${\bf k}$ vectors perpendicular to the defects. One gets qualitatively the
same results as in the case of point defects. 

In order to compare the effect of the different kinds of defects, we have
calculated the order parameter in all three cases (points, lines, and planes)
for the same defect strenght and the same defect concentration (i.e., the
extended defects are viewed as correlated accumulations of point defects with
the total number of -- pointlike -- defects held fixed). Therefore, the 
resulting differences solely originate in the different disorder
dimensionality. The one--dimensional soft sector was taken to be perpendicular 
to the defects in order to provide a meaningful comparison. The result is
depicted in Fig.~\ref{opdim}; it can be seen that the effect of the defects is 
not a monotonous function of their dimensionality, but depends on the strength 
of two competing effects. On the one hand, when the temperature is lowered
towards the phase transition temperature and the correlation length $\xi$
grows accordingly, the order parameter cluster around a $d'$--dimensional
defect grows proportional to $\xi^{d-d'}$. This effect renders low--dimensional
defects more effective in influencing bulk properties. On the other hand, the
system has a finite stiffness, characterized by the parameter $c$; and as a
system with uncorrelated defects is more inhomogeneous than one with the 
identical amount of correlated disorder, this effect favors high--dimensional
defects, because then the system stiffness can be more easily overcome by the
joint action of neighboring defects. With the specific numerical values we have
used, the line defects have only a tiny effect on the order parameter curve in 
comparison with the point defects. The effect of the planar defects lies in
between. We emphasize that this scenario could be different for other values of
the stiffness parameter $c$. Finally, we remark that if one performs the above
calculations for a system with extended defects, where the soft sector is not
perpendicular to the defects, additional angle dependences ensue, and one has 
to add an additional term of the form $cq^4$ to the functional (\ref{Fritzi}),
in order to correctly account for the stiffness, which tends to prevent the 
building--up of order parameter clusters.

\section{Summary and discussion}

In this paper we have studied the influence of point and extended defects on a
$d$--dimensional elastic system with an $m$--dimensional soft sector undergoing
an elastic phase transition of second order. We have calculated the
phonon--phonon correlation function in the high--temperature phase. At a
certain temperature $T_c(n_{\rm D}) \gg T_c^0$, an instability marking a
defect--induced phase transition may emerge, if the defect potentials are
sufficiently strong. Above $T_c(n_{\rm D})$ a dynamical central peak emerges in
the phonon correlation function, whose intensity grows as the temperature is
lowered towards $T_c(n_{\rm D})$. Contrary to the case of distortive phase
transitions \cite{Sch91}, the maximum of the central peak is exactly at
$\omega = 0$ for all temperatures and not at small but finite frequencies
\cite{Sch90}. This, however, does not imply that the acoustic impurity modes
are localized; at least for a simplifying one--dimensional single--defect model
no additional localized impurity modes appear, but the defect rather causes a
localized vibrational contribution to the propagating scattering states; for
the long--wavelength phonons this quasi--resonant vibration then condenses at
$T_c(n_{\rm D})$ and forms the local order parameter clusters \cite{Sch90}. The
dynamical central peak may be regarded as precursor of this phase transition;
its height also depends on the angle between the external wave vector ${\bf k}$
and the soft sector. The smaller this angle, the more pronounced is the central
peak.

In the low--temperature phase $T < T_c(n_{\rm D})$, where a finite, and be it
ever so small, order parameter exists, we have used a self--consistent
mean--field calculation in order to calculate the average order parameter, the
free energy and specific heat, and the phonon correlation function. The order
parameter sets in continuously at $T_c(n_{\rm D})$, remains very small in the
temperature range between $T_c(n_{\rm D})$ and $T_c^0$, and reaches appreciable
values only near $T_c^0$. In this way the order parameter curve resembles a
somewhat rounded curve of the pure system. Analogously, the temperature
dependence of the specific heat looks like the corresponding smeared--out curve
for the pure system. The jump at $T_c^0$ is rounded, and a minute jump at
$T_c(n_{\rm D})$ appears. Thus the phase transition of the perturbed system no
longer occurs at $T_c^0$ but at $T_c(n_{\rm D})$. However, the phase transition
temperature $T_c^0$ of the pure system remains important, as the remnants of
the pure transitions induce marked, but rounded maxima in quantities like the
order parameter susceptibility or the specific heat near $T_c^0$.

Having thus determined the order parameter, we were able to calculate the
phonon correlation function in the ordered phase. The dynamical central peak
disappears again as the temperature is lowered below $T_c(n_{\rm D})$. The
dependence of the central peak on the angle between the momentum transfer
vector ${\bf k}$ and the soft sector is very similar as above $T_c(n_{\rm D})$.
The density--density correlation function determining the cross section for
scattering experiments consists of three terms: first, the term describing
elastic Bragg scattering, second, a term corresponding to Huang scattering
caused by the spatially inhomogeneous order parameter configuration; this term
yields a contribution to elastic scattering, leading to a static central peak
with finite $k$ width. The third term finally describes inelastic scattering
and has been discussed along with the phonon correlation function.

We have also discussed the validity of our mean--field approach and provided an
estimate (in the isotropic case) for the ``true'' precursor $T_c$, defined as
the cluster ordering temperature $T_{\rm ord}$, which has to be distinguished
from the temperature $T_c(n_{\rm D})$ where localized, but fluctuating order
parameter clusters appear. Only below $T_{\rm ord}$, the previously independent
ordered regions form a collective state leading to a non--zero average order
parameter. Generally, cooperative behavior of the defects is to be expected
when the correlation length of the pure system $\xi$ becomes of the order of
the mean defect separation $r_{\rm D}$. Although the ensuing cluster ordering
temperature is considerably lower than $T_c(n_{\rm D})$, the qualitative
features of the present theory should remain largely unaffected, provided
$T_c(n_{\rm D})$ is replaced by $T_{\rm ord}$; i.e., the mean order parameter
appears rounded near $T_c^0$, while static order parameter susceptibility and
the specific heat display a strong but broadened maximum there. Furthermore, in
the anisotropic systems under consideration here, fluctuations may actually be
suppressed, rendering mean--field theory more reliable. Finally, the time scale
of the order parameter condensate fluctuations will diverge
$\propto (T - T_{\rm ord})^{-1}$ upon approaching $T_{\rm ord}$, which in
experiment would eventually render them indistinguishable from static
inhomogeneities, leading to quasi--elastic Bragg and Huang scattering peaks. We
have also investigated the case of parallel extended defects (lines and
planes), and found essentially the same features \cite{Bul93}.

These considerations led us to the qualitative phase diagram displayed in
Fig.7. For very tiny disorder concentrations, the picture of isolated defects
applies. Although preordered clusters may form considerably above $T_c^0$, the
ensuing condensates fluctuate independently and do not form a state with
nonzero average order parameter. The cluster reorientation rate will become
very low as $T_c^0$ is approached, and coupling of these slow modes to the soft
(acoustic) phonons will then lead to a dynamical central peak, see
Ref.~\cite{Sch74}. For higher, but still small defect concentrations, the
clusters emerging at $T_c(n_{rm D})$ will form a state with preordered defect
regions and finite, but small average order parameter below $T_{\rm ord}$.
This phase transition leads to discontinuities in thermodynamic quantities,
like the specific heat or the static susceptibility, which are, however,
probably unnoticeably small in experiment. On the other hand, upon approaching
$T_c^0$, the phase transition temperature of the pure crystal, the bulk system
orders; due to the influence of the preordered defect regions, the transition
temperature will be slightly higher than $T_c^0$, and the formerly sharp
singularities of the mean order parameter, specific heat, and static
susceptibility appear characteristically rounded. The above described theory
applies precisely to this concentration range (for very strong disorder the
single--site approximation breaks down). The central peak phenomenon is thus
explained by a combination of elastic Bragg peaks of the low--temperature
phase and static Huang scattering with finite width in $q$ space.

However, a different scenario is also conceivable, namely that as the
cluster reorientation times become very long, the different, still independent
condensates freeze in with spatially fluctuating orientations. The ensuing
configuration would constitute a metastable state which is separated from the
true ground state by high free energy barriers $\Delta$. The typical flip rate
would then be proportional to $\exp(-\Delta / k_{\rm B} T)$; therefore, at low
temperatures the true thermodynamic ground state may not be reached. In the
spirit of the discussion in Sec.V one could possibly map this problem onto a
random--field Ising model (for recent reviews, see, e.g. Ref.~\cite{Nat88}),
the lower critical dimension of which is $d_l = 2$, and therefore long--range
order is not destroyed in three dimensions.

At last, we would like to contrast our picture with that of "glassy" systems.
Although some of the features of the dynamical structure factor in glasses are
at least qualitatively similar, e.g., an elastic peak with finite $q$ width 
appears, and the static susceptibility and specific heat may display
characteristically rounded and broadened maxima, there are important
differences. First, the order parameter of the pure crystal $\bar{\nu}$ would
not constitute an appropriate order parameter for such a disordered, glassy
system. Second, the character of the phase transition should be entirely
different, and in fact lead to experimentally distinguishable behavior. The
scenario described here is a genuine second--order phase transition, though
induced by disorder (which locally softens the system); i.e.: critical
phenomena are confined to regions very close to $T_c(n_{\rm D})$ (or
$T_{\rm ord}$), and to wave vectors in the soft sector with
${\bf p} \approx 0$. The freezing--in into a glassy state, on the other hand,
would have to be described by an ergodicity--breaking (Edwards--Anderson) order
parameter, and should actually be rather insensitive to $k$. Such a glass
instability occurs, e.g., in orientational glasses \cite{Mic87}, and possibly
in relaxor ferroelectrics \cite{Vie92}. We finally remark that for the case of
first--order martensitic transformations, a model with disorder of the random
$T_c$ type has been proposed, which can then be mapped onto a spin glass, and
the ensuing glassy features were suggested to explain the prominent tweed
microstructure found in these materials, as well as the central peak phenomenon
there \cite{Kar91}.

\acknowledgments

U.C.T. acknowledges support from the Deutsche Forschungsgemeinschaft (DFG)
under Contract Ta. 177/1-2.

$^*$ Present address: University of Oxford, Department
of Physics -- Theoretical Physics, 1 Keble Road, Oxford OX1 3NP, U.K.

\end{multicols}

\begin{table}[b]
\caption{Ginzburg-Landau parameters, as used in the figures, if not specified
         otherwise [$a=a'(T-T_c^0)$].}
\begin{tabular}{l}
   $T_c^0=45$ K\\
   $a'= 1.0771 \cdot 10^{-12} \mbox{erg K}^{-1}$\\
   $d = 1.1363 \cdot 10^{-8}  \mbox{erg}$\\
   $M=1.314 \cdot 10^{-21} \mbox{g}$\\
   $b=1.570 \cdot 10^{-10} \mbox{erg}$\\
   $c= 5 \cdot 10^{-26} \mbox{erg cm}^2$\\
   $N/V= 6.751 \cdot 10^{21} \mbox{cm}^{-3}$\\
   $\lambda= 3.0994 \cdot 10^{-11} \mbox{erg}$\\
   $D=\tilde{D}=1 \cdot 10^{-3} \mbox{cm}^2 \mbox{s}^{-1}$\\
   $k= \zeta\cdot \sqrt{2}\cdot a^{*}$\\
   $a^{*}= 2 \pi / a_0= 1.189\cdot 10^{8} \mbox{cm}^{-1}$\\
   $T_c(n_{\rm D})= 65 \mbox{K} $\\
   $n_{\rm D}= 1 \cdot 10^{-5}  $\\
   $\lambda= 3.0994 \cdot 10^{-11} \mbox{erg}$\\
\end{tabular}
\end{table}

\begin{figure}[p]
  \centerline{\rotate[r]{\epsfysize=4.5in \epsffile{pic12.ps.bb}}}
  \label{MECObild12}
\end{figure}

\begin{figure}[p]
  \centerline{\rotate[r]{\epsfysize=4.5in \epsffile{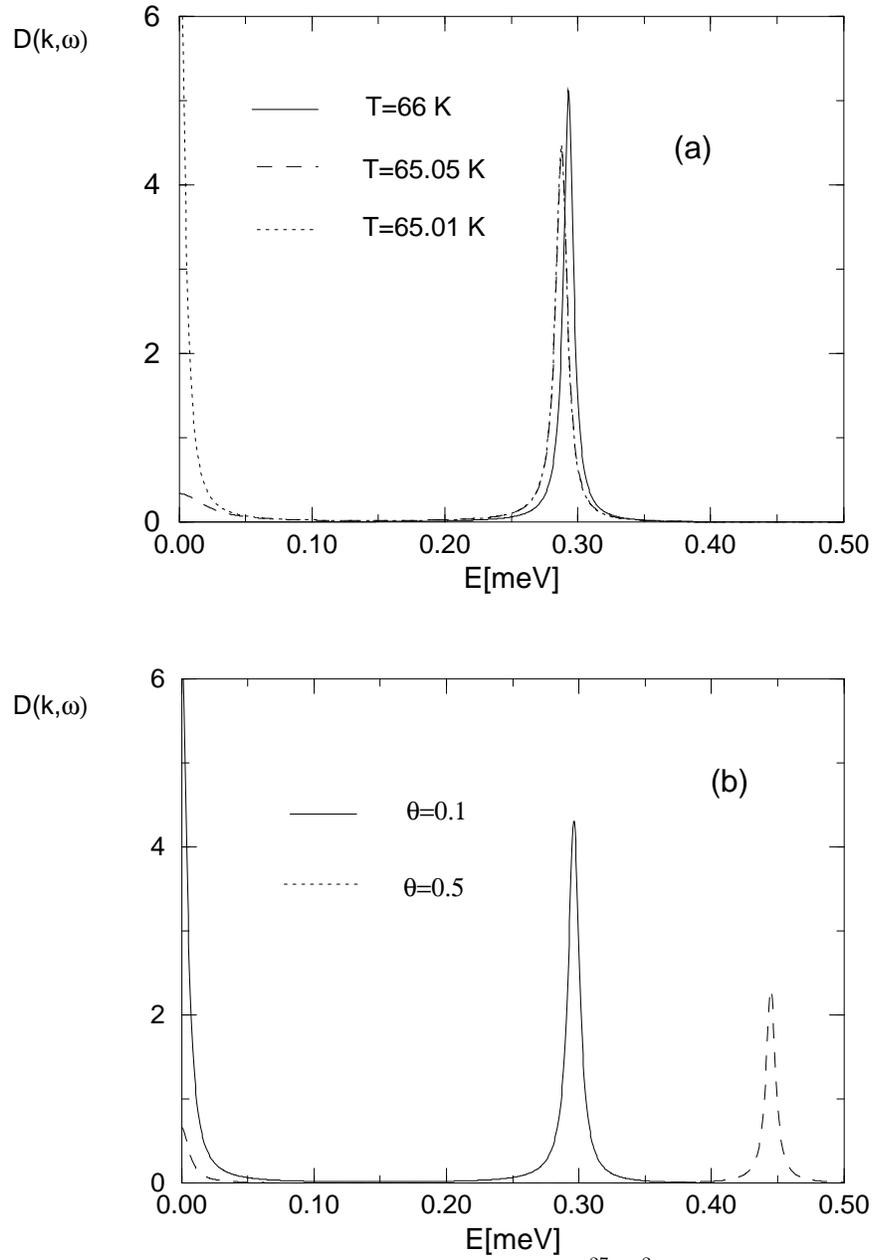}}}
  \caption{Phonon correlation function
  $D({\bf k},\omega) = 2 k_{\rm B}T \rm{Im} G({\bf k},\omega) / \omega$ 
  [in $10^{-27} {\rm cm^2 s}$] vs. phonon energy [in meV] for different
  temperatures and fixed angle $\theta=0$ (a), and for fixed temperature
  $T=65.01$ K with different angles $\theta$ (b);
  $\zeta = k a_0/ 2^{3/2} \pi = 0.02$ was used.}
\label{MECObild11}
\end{figure}

\begin{figure}[p]
  \centerline{\rotate[r]{\epsfysize=4.5in \epsffile{pic21.ps.bb}}}
  \label{MECObild21}
\end{figure}

\begin{figure}[p]
  \centerline{\rotate[r]{\epsfysize=4.5in \epsffile{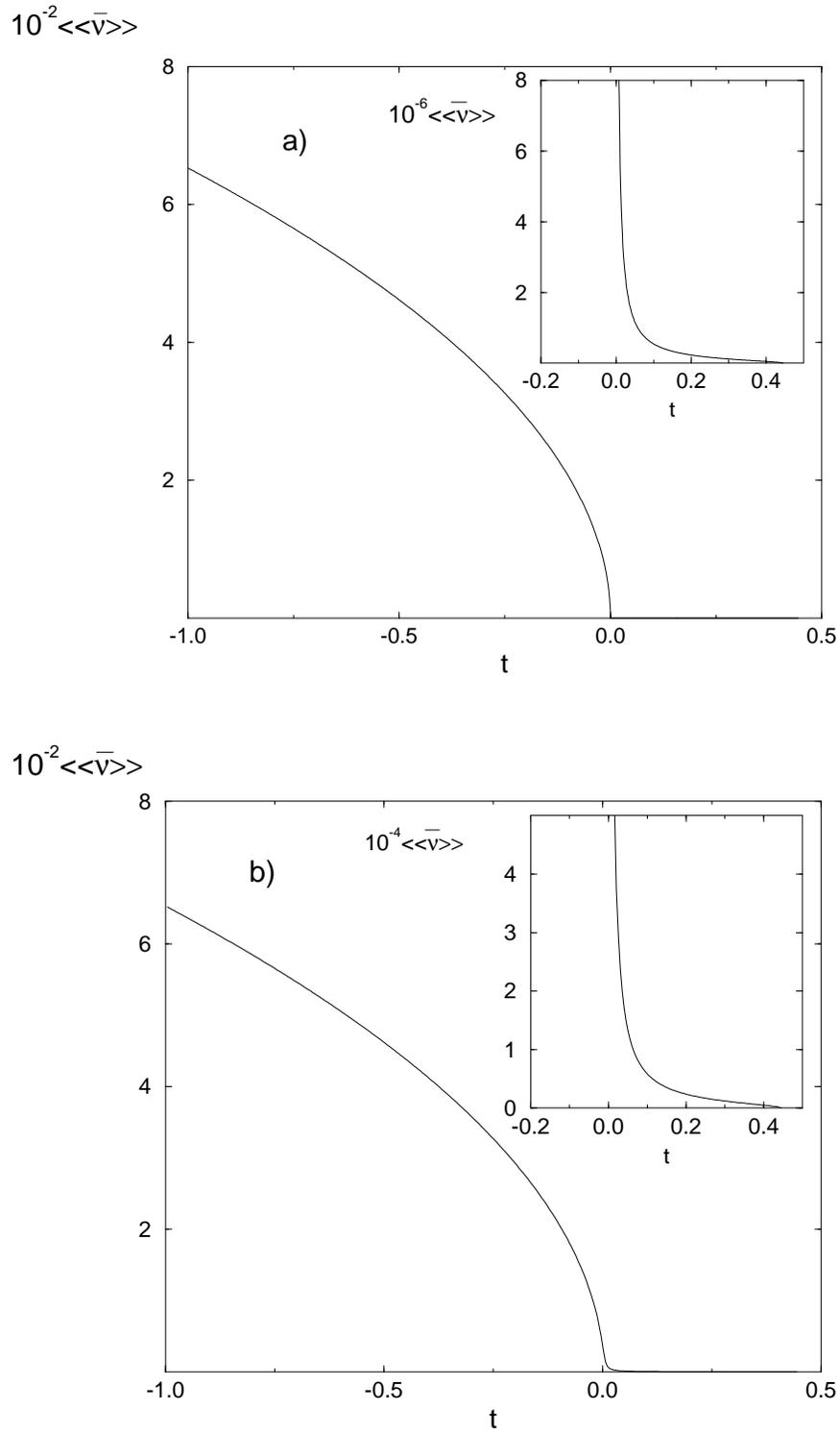}}}
   \caption{Average order parameter $\langle\langle {\bar \nu} \rangle\rangle$ 
  vs. reduced  temperature $t$ for $n_{\rm D}=10^{-5}$, $\lambda=3.0994 \cdot 
  10^{-11} \mbox{erg}$ (a), and $n_{\rm D}=10^{-3}$,
  $\lambda= 3.0949 \cdot 10^{-11} \mbox{erg} $ (b). The temperature range near
  $t_c(n_{\rm D}) = 0.444$ is displayed in the insets. Note the different
  scales of the two insets; the disorder strength was adjusted to yield the
  same $T_c(n_{\rm D})$ in both cases (a) and (b).}
\label{MECObild22}
\end{figure}

\begin{figure}[p]
  \centerline{\rotate[r]{\epsfysize=4.5in \epsffile{pic31.ps.bb}}}
   \label{MECObild31}
\end{figure}

\begin{figure}[p]
  \centerline{\rotate[r]{\epsfysize=4.5in \epsffile{pic32.ps.bb}}}
   \caption{Specific heat $C_v$ vs. reduced temperature $t$ for $n_{\rm
  D}=10^{-5}$ (a) and $n_{\rm D}=10^{-3}$ (b). The defect--induced temperature
  has been adjusted to $t_c(n_{\rm D}) = 0.444$ in both cases by changing the
  defect potential strength accordingly, see Fig.2. The temperature range near
  $t_c(n_{\rm D})$ is displayed in the insets; note the different scales.} 
\label{MECObild3}
\end{figure}

\begin{figure}[p]
   \centerline{\rotate[r]{\epsfysize=4.5in \epsffile{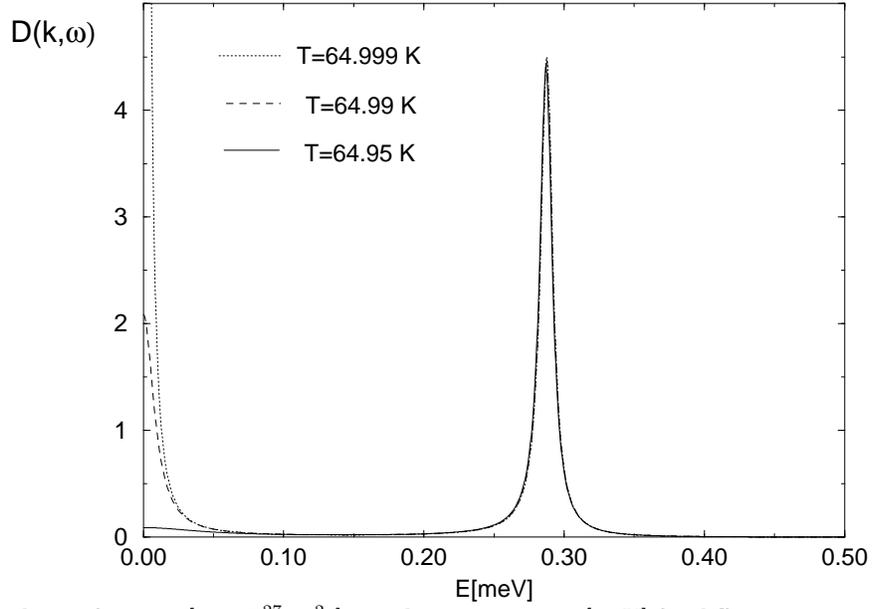}}}
   \caption{Phonon correlation function [in $10^{-27} {\rm cm^2 s}$] vs.
  phonon energy in [meV] for different temperatures below $T_c(n_{\rm D})$ at 
  fixed angle $\theta=0$. As before, $t_c(n_{\rm D})=0.444$, and
  $\zeta = k a_0/ 2^{3/2} \pi = 0.02$ was used.}
\label{MECObild4}
\end{figure}

\begin{figure}[p]
  \centerline{\rotate[r]{\epsfysize=4.5in \epsffile{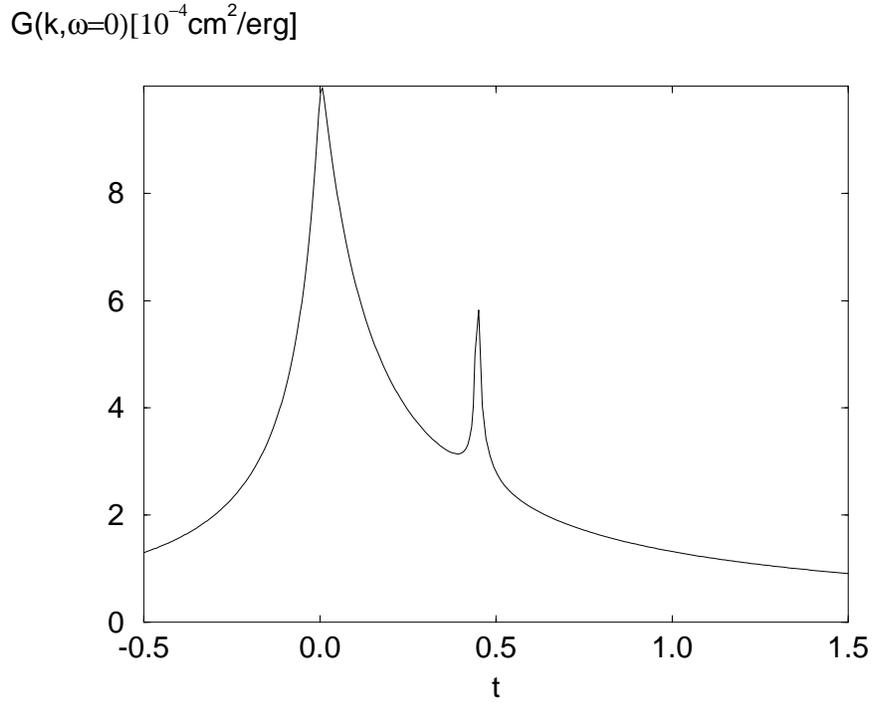}}}
  \caption{Static phonon susceptibility (inverse elastic constant) $G({\bf k})$
  [in $10^{-4} {\rm cm^{2}} \mbox{erg}^{-1}$] vs. temperature at fixed angle
  $\theta=0$. 9
  $t_c(n_{\rm D})=0.444$, $\zeta = k a_0/ 2^{3/2} \pi = 0.07$.}
\end{figure}

\begin{figure}[p]
  \centerline{\rotate[r]{\epsfysize=4.5in \epsffile{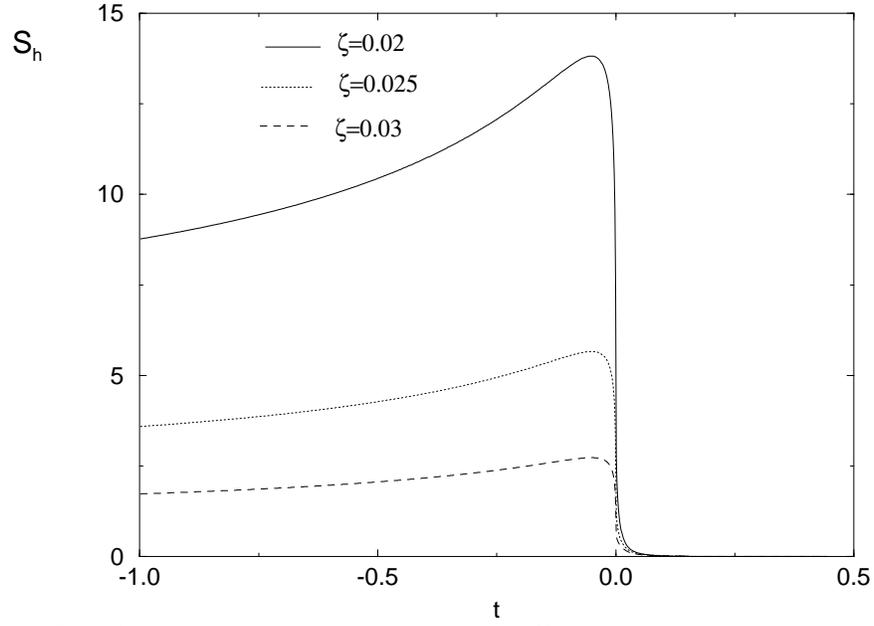}}}
  \caption{The term $\langle\langle\bar{\nu}\rangle\rangle ^2 
  \tilde{G}_0({\bf k})^2 (a + d A^2)^2$, denoted by $S_h$ [in units $10^{-11}$]
  vs. reduced temperature $t$ for different wavevectors. The wave vector
  ${\bf k}$ lies in the soft sector.}
\label{MECObild5}
\end{figure}

\begin{figure}[p]
  \centerline{\rotate[r]{\epsfysize=4.5in \epsffile{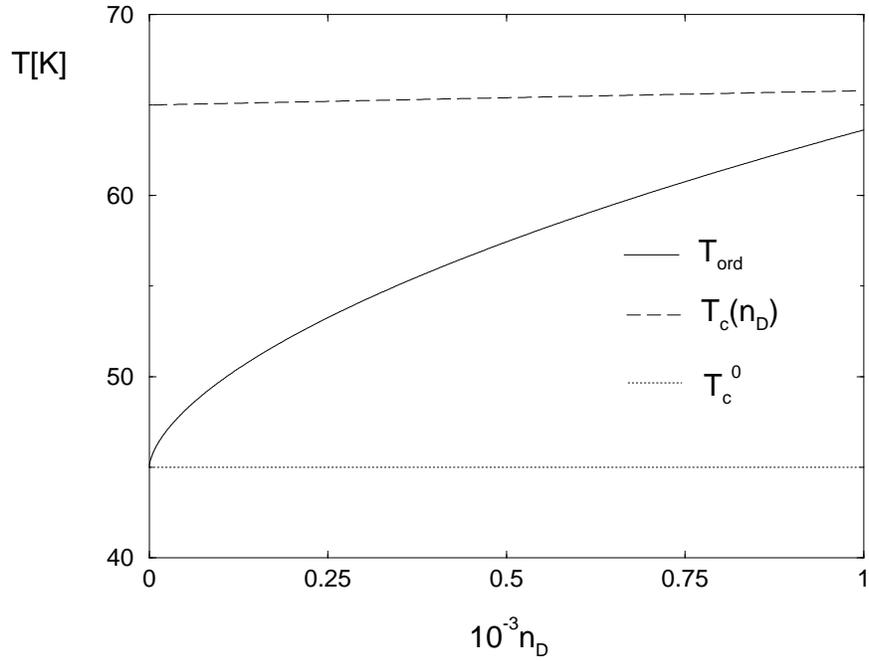}}}
  \caption{Cluster ordering (precursor) temperature $T_{\rm ord}$ and
  mean--field (local) transition temperature $T_c(n_{\rm D})$ [in K] as
  function of the defect concentration $n_{\rm D}$; the numerical values of
  Table I were used here.}  
\end{figure}

\begin{figure}[p]
  \centerline{\rotate[r]{\epsfysize=4.5in \epsffile{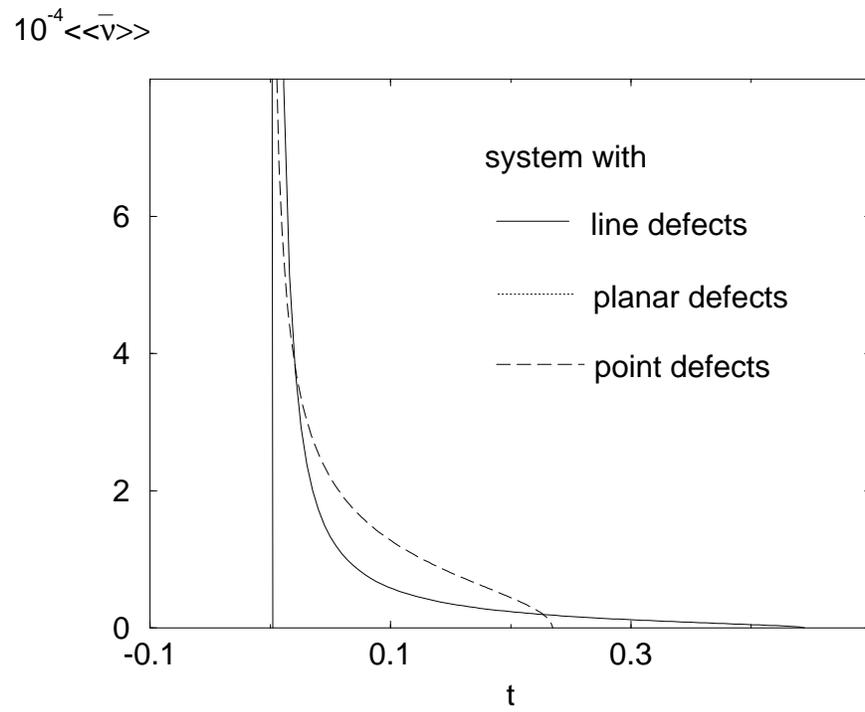}}}
  \caption{Order parameter vs. reduced temperature for systems with different
  types of defects; the numerical values of
  Table I and $n_{\rm D}=10^{-3}$ were used here.}  
  \label{opdim}
\end{figure}

\end{document}